\documentclass[aps,pre,twocolumn,showpacs,superscriptaddress]{revtex4}

\usepackage{amssymb}
\usepackage{amsmath}
\usepackage[british]{babel}
\usepackage{graphicx}
\usepackage{color}

\begin{document}

\title{Ergodicity breaking and particle spreading in noisy heterogeneous diffusion
processes}

\author{Andrey G. Cherstvy}
\affiliation{Institute for Physics \& Astronomy, University of Potsdam,
14476 Potsdam-Golm, Germany}
\email{a.cherstvy@gmail.com}
\author{Ralf Metzler}
\affiliation{Institute for Physics \& Astronomy, University of Potsdam,
14476 Potsdam-Golm, Germany}
\affiliation{Department of Physics, Tampere University of Technology, 33101 Tampere, Finland}
\email{rmetzler@uni-potsdam.de}

\date{\today}

\begin{abstract}
We study noisy heterogeneous diffusion processes with a position dependent
diffusivity of the form $D(x)\sim D_0|x|^\alpha$ in the presence of annealed
and quenched disorder of the environment, corresponding to an effective variation
of the exponent $\alpha$ in time and space. In the case of annealed disorder, for
which effectively $\alpha=\alpha(t)$ we show how the long time scaling of the
ensemble mean squared displacement (MSD) and the amplitude variation of individual
realizations of the time averaged MSD are affected by the disorder strength. For
the case of quenched disorder, the long time behavior becomes effectively
Brownian after a number of jumps between the domains of a stratified medium. In
the latter situation the averages are taken over both an ensemble of particles and
different realizations of the disorder. As physical observables we analyze in
detail the ensemble and time averaged MSDs, the ergodicity breaking parameter, and
higher order moments of the time averages.
\end{abstract}


\maketitle

\section{Introduction}

The motion of individual molecules and submicron tracer particles of different
sizes in the cytoplasm of living biological cells \cite{cells}, in artificially
crowded environments in vitro \cite{invitro}, in glass-like systems \cite{glass},
or in large scale in silico studies of membrane structures \cite{membranes} was
shown to follow the anomalous diffusion law
\begin{equation}
\langle x^2(t)\rangle\simeq t^\beta,
\label{msd}
\end{equation}
with the subdiffusive diffusion exponent mostly in the range $\beta=0.4\dots0.9$
\cite{report,franosch}. A number of mathematical models of different kinds were
proposed to unveil the properties of anomalous diffusion phenomena embodied in
the mean squared displacement (MSD) in Eq.~(\ref{msd}) \cite{pccp}. In most of
these models the properties of the stochastic process are homogeneous in space.
Especially for smaller tracers, which may cover longer
distances within the measurement time, or for techniques allowing for full maps
of local diffusivities, it turns out that the diffusion coefficient becomes a
function of the local tracer position. For both eukaryotic \cite{lang11} and
prokaryotic \cite{elf11} cells such local diffusivity maps indeed show significant
variations. The motion of tracer particles through space may also be impeded by
caging effects when
the size of the particle is comparable to the local mesh size in structured
environments \cite{wong,aljaz}. In such cases the tracer diffusion becomes
characterized by a non-uniform, position-dependent diffusivity $D(x)$. Similarly,
spatially varying transport characteristics are ubiquitous in contaminant
dispersion in subsurface water aquifers \cite{hydro}.

In the field of stochastic dynamics anomalous diffusion in spatially random media,
disordered
energy landscapes, weakly chaotic systems, and dynamic maps received considerable
attention \cite{bouchaud,havlin,ledoussal-diff,papa95-random,kehr-review,
chaotic-diff,trap-models,maps}. More specifically, anomalous diffusion due to
micro-domains was investigated \cite{dagdug14BJ} and the influence of environmental
Gaussian noise on diffusive particle trajectories in disordered systems was
studied \cite{nCTRW}. Moreover, deviations from normal diffusion due to quenched
and annealed disorder of the medium diffusivity received renewed interest
\cite{spain14-pd,slater14}. In such studies one is mainly interested in the
quantitative behavior of the particle MSD (\ref{msd}) as
well as the ergodic properties of the system: is the information from time averages
of physical observables typically garnered as time series by modern particle
tracking assays equivalent to those of the corresponding ensemble averages known
from the theoretical models? It turns out that a large variety of anomalous
diffusion processes involve weak ergodicity breaking \cite{web,web1,pccp,pt,igor,
pccp11}, the disparity between (long) time averages and ensemble averages of physical
observables such as the MSD, and that in those cases the Khintchine theorem needs
to be substituted by generalized versions \cite{khinchin,pnas}.

\begin{figure}
\includegraphics[width=8cm]{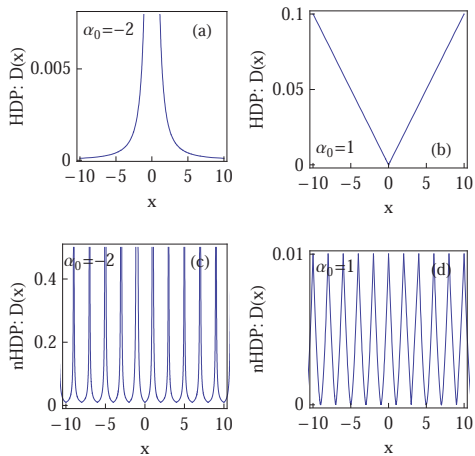}
\caption{Particle diffusivity for heterogeneous diffusion processes: (a) and (b)
for, respectively, $\alpha_0=-2$ and $\alpha_0=1$. Panels (c) and (d) show the
diffusivity of HDPs with quenched disorder for the same values of $\alpha_0$
and for the parameters $\sigma^2=0.25$ and $2\delta x=2$ (see text for details). 
Slight variations of the diffusivity in panels (c) and (d) indicate the external
noise superimposed in the process.}
\label{fig-diff-coeff}\end{figure}

Here, we study the dynamics and the ergodic properties of heterogeneous diffusion
processes (HDPs) with position dependent diffusivity $D(x)$, in the presence of
piece-wise deterministic quenched and annealed disorder. More specifically, we
generalize HDPs with power-law diffusivity
\begin{equation}
D(x)=D_0|x|^{\alpha_0},
\label{eq-d-alfa}
\end{equation}
for which the anomalous diffusion exponent of the MSD assumes the form
\cite{fulinski,hdp,hdp1,hdp2,hdp3}
\begin{equation}
\label{hdp_exp}
\beta=\frac{2}{2-\alpha_0}.
\end{equation}
The physical dimension of the coefficient $D_0$ in Eq.~(\ref{eq-d-alfa}) is $[D_0]=
\mathrm{cm}^{2-\alpha_0}\mathrm{sec}^{-1}$. The exponent (\ref{hdp_exp}) designates
subdiffusion for $\alpha_0<0$ and superdiffusion for $0<\alpha_0$ \cite{fulinski,
hdp,hdp1,hdp2,hdp3}. The profiles of the diffusivity for these cases are shown in
Fig.~\ref{fig-diff-coeff}a,b. HDPs are weakly non-ergodic and ageing, that is, their
dynamics depends explicitly on the time gap between original initiation of the
system and start of the measurement \cite{fulinski,hdp,hdp1,hdp2,hdp3}. We note
that the ageing properties of HDPs \cite{hdp2} embodied in the ensemble and time
averaged MSDs are in fact similar to those of subdiffusive continuous time random
walks \cite{johannes} and scaled Brownian motion \cite{hadiseh}.

In the following we unravel how the additional disorder in the system modifies the
diffusive and ergodic properties of HDPs. We compute the scaling laws for the
ensemble and time averaged MSDs as well as the amplitude spread of individual
realizations of the process. The article is structured as follows. In Sec.~II we
define the observables, that we will analyze. Sec.~III specifies the model and
its implementation in the simulations. In Sec.~\ref{sec-annealed} we then study
HDPs with annealed disorder, followed by the scenario with quenched disorder in
Sec.~\ref{sec-quenched}. Sec.~VI concludes this work.

\section{Observables}

The central quantity in the study of stochastic processes is the ensemble
averaged MSD
\begin{equation}
\langle x^2(t)\rangle=\int_{-\infty}^{\infty}x^2P(x,t)dx,
\end{equation}
calculated as the spatial average of $x^2$ over the probability density function
$P(x,t)$ to find the particle at position $x$ at given time $t$ \cite{vankampen}.
However, when individual time series $x(t)$ of the particle position are measured
in experiments or simulations, the typical quantity studied is the time averaged MSD
\begin{equation}
\label{eq-TAMSD}
\overline{\delta^2(\Delta)}=\frac{1}{T-\Delta}\int_0^{T-\Delta}\Big[x(t+\Delta)-x(t)
\Big]^2 dt.
\end{equation}
Here $\Delta$ is the lag time and $T$ is the measurement time (length) of the
trajectory $x(t)$ \cite{pccp,pt,pccp11}. Often, also the additional average
\begin{equation}
\left<\overline{\delta^2(\Delta)}\right>=\frac{1}{N}\sum_{i=1}^N\overline{\delta^2_i
(\Delta)}
\end{equation}
of the time averaged MSD over $N$ individual trajectories is taken \cite{pccp,pt,
pccp11}. A process is called ergodic when we observe the equality
\begin{equation}
\label{ergo}
\langle x^2(\Delta)\rangle=\lim_{\Delta/T\to0}\overline{\delta^2(\Delta)}.
\end{equation}
Examples for ergodic processes are Brownian motion \cite{pccp,pccp11,igor,pt}
as well as anomalous diffusion processes with MSD (\ref{msd}) given by random
walks on fractals \cite{yasmine} and processes driven by fractional Gaussian noise
\cite{deng,pre,goychuk}. Once a process is non-stationary, the equality (\ref{ergo})
is violated, the phenomenon of weak ergodicity breaking \cite{web,web1,pt,pccp,
pccp11,igor}. A whole range of anomalous diffusion processes with power-law MSD
(\ref{msd}) belongs to this class and specifically exhibits the linear lag time
dependence
\begin{equation}
\left<\overline{\delta^2(\Delta)}\right>\simeq\frac{\Delta}{T^{1-\beta}}
\label{eq-tamsd-hdp}
\end{equation}
of the time averaged MSD \cite{pccp}. As examples we mention
continuous time random walk processes with scale free distributions of waiting
times \cite{web,web1,pt,pccp,pccp11,igor,pnas}, correlated continuous time random
walks \cite{cctrw}, as well as diffusion processes with space \cite{fulinski,hdp,
hdp1,hdp2,hdp3} and time \cite{fulinski,thiel,sbm,hadiseh} dependent diffusion
coefficients
and their combinations \cite{gdp}. We also mention ultraslow diffusion processes
with a logarithmic form for $\langle x^2(t)\rangle$ and linear lag time dependence
(\ref{eq-tamsd-hdp}) of the time averaged MSD \cite{ultraslow} as well as the
ultraweak ergodicity breaking of superdiffusive L{\'e}vy walks \cite{walks}.

For finite measurement time even ergodic processes exhibit a statistical scatter
of the amplitude of time averaged observables. This irreproducibility for the
case of the time averaged MSD $\overline{\delta^2(\Delta)}$ can be quantified in
terms of the distribution $\phi(\xi)$ as function of the dimensionless variable
\cite{pccp,pt,pccp11,web1}
\begin{equation}
\xi=\frac{\overline{\delta^2}}{\left<\overline{\delta^2}\right>}. 
\end{equation}
The variance of $\phi(\xi)$ is quantified in terms of the ergodicity breaking
parameter \cite{web1,pccp,pt,pccp11}
\begin{equation}
\textrm{EB}(\Delta)=\left<\xi^2(\Delta)\right>-\left<\xi(\Delta)\right>^2\equiv
\left<\xi^2\right>-1.
\label{eq-eb-via-xi}
\end{equation}
For Brownian motion the behavior of the ergodicity breaking parameter at
$\Delta/T\to0$ is
\begin{equation}
\mathrm{EB}_{\text{BM}}(\Delta)=\frac{4\Delta}{3T}.
\label{eq-eb-bm}
\end{equation}
Continuous time random walk processes with scale free waiting time distribution
have a finite value for $\textrm{EB}$ even in the limit $\Delta/T=0$ \cite{web1},
similar to HDPs \cite{hdp,hdp1,hdp2,hdp3}, while for scaled Brownian motion the
ergodicity breaking parameter approaches zero in this limit \cite{thiel,sbm}.

For reference in what follows we also mention that the probability density function
of HDPs obeys has the exponential form \cite{hdp}
\begin{equation}
\label{eq-hdp-pdf}
P(x,t)=\frac{|x|^{-2/\alpha_0}}{\sqrt{4\pi D_0t}}\exp\left(-\frac{|x|^{
2-\alpha_0}}{(2-\alpha_0)^2 D_0 t}\right)
\end{equation}
which is a stretched (compressed) Gaussian for superdiffusive (subdiffusive) HDPs
with $0<\alpha_0<2$ ($\alpha_0<0$). Note that, respectively, the shape
(\ref{eq-hdp-pdf}) has a distinct cusp at the origin or is bimodal with $P(0,t)=0$
\cite{hdp}.

\section{Model}

We employ the same tested stochastic algorithm for the Markovian HDPs as developed
in Refs.~\cite{hdp,hdp1,hdp2,hdp3}, based on the one-dimensional Langevin equation
for the particle displacement $x(t)$ with the position dependent diffusivity $D(x)$,
\begin{equation}
\frac{dx(t)}{dt}=\sqrt{2D(x)}\times\zeta(t).
\label{eq-langevin}
\end{equation}
The process is driven by the white Gaussian noise $\zeta(t)$ with covariance
$\langle\zeta(t)\zeta(t')\rangle=\delta(t-t')$ and zero mean $\langle\zeta(t)
\rangle=0$. We interpret Eq.~(\ref{eq-langevin}) in the Stratonovich sense
leading to the following implicit mid-point iterative scheme: at step $i+1$ the
particle position is
\begin{equation}
\label{eq-simul-scheme}
x_{i+1}-x_i=\sqrt{2D\left(\frac{x_{i+1}+x_i}{2}\right)}\times(y_{i+1}-y_{i}),
\end{equation}
where the increments $(y_{i+1}-y_i)$ of the Wiener process represent a
$\delta$-correlated Gaussian noise with unit variance and zero mean. Unit time
intervals separate consecutive iteration steps. Below we simulate three values
for the exponent $\alpha_0$, corresponding to $\beta=1/2$ (subdiffusive MSD), 
$\beta=0$ (Brownian motion), and $\beta=2$ (superdiffusive MSD). For standard
HDPs these cases were analyzed by us in Refs.~\cite{hdp,hdp1,hdp2,hdp3}. To avoid
divergencies of the particle motion we regularize the diffusivity at $x=0$ by
addition of a small constant, namely $D(x)=D_0(|x|^\alpha+D_{\mathrm{off}})$ where 
$D_{\mathrm{off}}=10^{-3}$ and $D_0=10^{-2}$ for all results shown below. This
choice does not affect the quality of the studied scaling laws \cite{hdp}.

To examine the effect of additional noise due to the environment we implement a
Gaussian distribution of the scaling exponent of the diffusivity with the mean
$\alpha_0$,
\begin{equation}
\label{eq-palfa}
p(\alpha)=\frac{1}{\sqrt{2\pi\sigma^2}}\exp\left(-\frac{(\alpha-\alpha_0)^2}{2
\sigma^2}\right).
\end{equation}
Generally, the distribution $p(\alpha)$ may be asymmetric, but we restrict our
discussion to symmetric forms. We consider two versions of this additional
disorder corresponding to the annealed and quenched limits for the variation of
$\alpha$. In the annealed case of noisy HDPs, the properties of the environment
change rapidly in time compared to time scales of the particle motion. Physically, 
such noise may be due to the imprecision of the experimental setup or because of
additional thermal agitation in the system. The diffusing particle thus visits
regions in space with different local exponents $\alpha$. In this scheme the
particle diffusivity at position $x$ fluctuates in time, and the value of the
diffusivity will be different each time the particle revisits the same position
$x$. In this annealed case large diffusivity variations occur in the entire space.
 
In superdiffusive HDPs distant particle excursions take place due to the growth of
$D(x)$ away from the origin and the associated acceleration of the motion, while
for subdiffusive
HDPs the walker is increasingly trapped in the low-diffusivity regions at larger
value of the position $|x|$ \cite{hdp,hdp1,hdp2,hdp3}. With increasing strength
$\sigma^2$ of the annealed noise given by the distribution (\ref{eq-palfa}) the
excursions of the particles in both superdiffusive and subdiffusive cases become
more erratic as time evolves. The time interval $\delta t$ during which the walker
has a given HDP exponent $\alpha_i$ obviously affects the properties of noisy HDPs. 
These time spans $\delta t$ are here taken to be uniformly distributed. To simulate
annealed noisy HDPs we use Eq.~(\ref{eq-palfa}) with varying $\sigma^2$. The
particle performs jumps with a given scaling exponent for the time interval $\delta
t$, after which a new exponent is chosen from the distribution (\ref{eq-palfa}), and
so on. The particle displacement $x_i$ during the time span $(t_i,t_i+\delta t)$ 
with HDP exponent $\alpha_i$ is the starting condition for the next time interval. 
Shorter $\delta t$ intervals imply more erratic motion, as shown below.

For noisy HDPs in the presence of quenched disorder, the profile of the particle 
diffusivity is hard-wired into the environment. We choose a static periodic
arrangement of domains as shown in Fig.~\ref{fig-diff-coeff}c,d. In each domain
the exponent $\alpha$ is drawn from $p(\alpha)$ and the particle performs a regular
HDP. The midpoint of each domain is chosen as the origin in the local HDP coordinate
system, that is, locally the functional shape of $D(x)$ is centered and decays or
increases with the local scaling exponent $\alpha$, as exemplified in
Fig.~\ref{fig-diff-coeff}c,d. The period $\delta x$ for the stratification of the
environment plays the role of a switching mechanism affecting the system dynamics. 
At the boundary of the domains the diffusivity and its derivative in general acquire
jumps. Physically, the latter occurs in the presence of some walls, cages, etc.

We simulate quenched noisy HDPs as follows. The entire space is stratified into domains
of width $2\delta x$, and the local HDP exponent is chosen from the distribution
(\ref{eq-palfa}). The length $\delta x$ is a vital parameter of quenched noisy HDPs. 
The particle performs an HDP random walk in each space domain with $D(x,\alpha)$ 
and it hops to a neighboring domain once the domain boundary is reached. The
centers of the domains are computed from the particle position $x_i$ as 
\begin{equation}
\label{eq-xc-defin}
x_{c,i}=2(\delta x)\textrm{int}\left[\frac{x_i}{2\delta x}\right]+\textrm{sign}
[x_i](\delta x),
\end{equation}
see Fig.~\ref{fig-diff-coeff}c,d. Here $\mathrm{int}[x]$ denotes the integer part
of the argument, and an additional $\delta x$ shift is used for convenience. The
starting position of the particle is near the center of the first domain, at $x(0)
=0.1+x_{c,1}$. The subsequent position $x_{i+1}$ is evaluated from $x_i$ with the
local exponent $\alpha_i$ according to Eq.~(\ref{eq-simul-scheme}), that is,
\begin{eqnarray}
\nonumber
x_{i+1}-x_i&=&\sqrt{2D_0\left(\left|\frac{x_{i+1}+x_i}{2}-x_{c,i}\right|^{\alpha_i}+
D_{\mathrm{off}}\right)}\\
&&\times(y_{i+1}-y_{i}).
\end{eqnarray} 
We vary the width of $p(\alpha)$ and the mean value of the scaling exponent $\alpha
_0$. Shorter periodicities $\delta x$ are equivalent to stronger external noise, as
shown below. We note here that for subdiffusive HDPs the centers of the domains
$x_{c,i}$ correspond to the regions of maximal diffusivity, while for superdiffusive
HDPs these are the spots of the lowest diffusivity \cite{hdp,hdp1}.

\section{Noisy HDPs with Annealed Disorder}
\label{sec-annealed}

\begin{figure*}
\includegraphics[width=16cm]{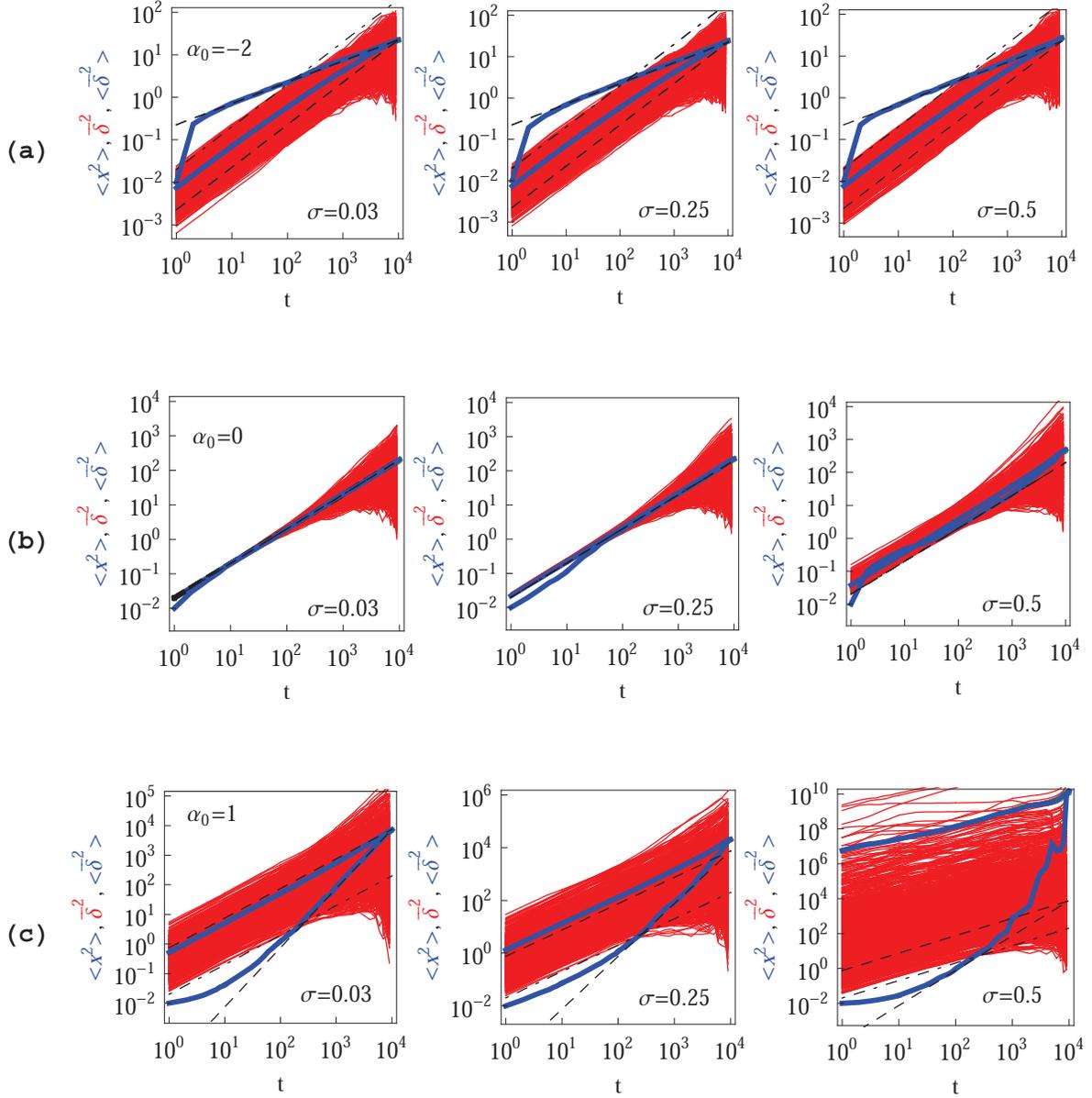}
\caption{Ensemble and time averaged MSDs $\langle x^2(t)$ and $\left<\overline{
\delta^2(\Delta)}\right>$ (thick blue curves) as well as individual time averaged
MSDs $\overline{\delta^2}$ (red curves) for annealed noisy HDPs. Parameters: the
values of $\alpha_0$ and its variance are indicated in the plots, the trace length
is $T=10^4$, and the number of sampled traces is $N=10^3$. The initial position
is $x_0=x(t=0)=0.1$. The top panels correspond to the noisy subdiffusive case, the
middle panel represents noisy Brownian motion, and the bottom panels are the case
of superdiffusive noisy HDPs. The asymptotes (\ref{msd}) and (\ref{eq-tamsd-hdp})
for the ensemble and time averaged MSDs of standard HDPs are shown as the dashed
curves. The Brownian asymptote $\langle x^2(t)\rangle=2D_0t$ is the dashed-dotted
line.}
\label{fig-ann-tamsd}
\end{figure*}

\begin{figure*}
\includegraphics[width=18cm]{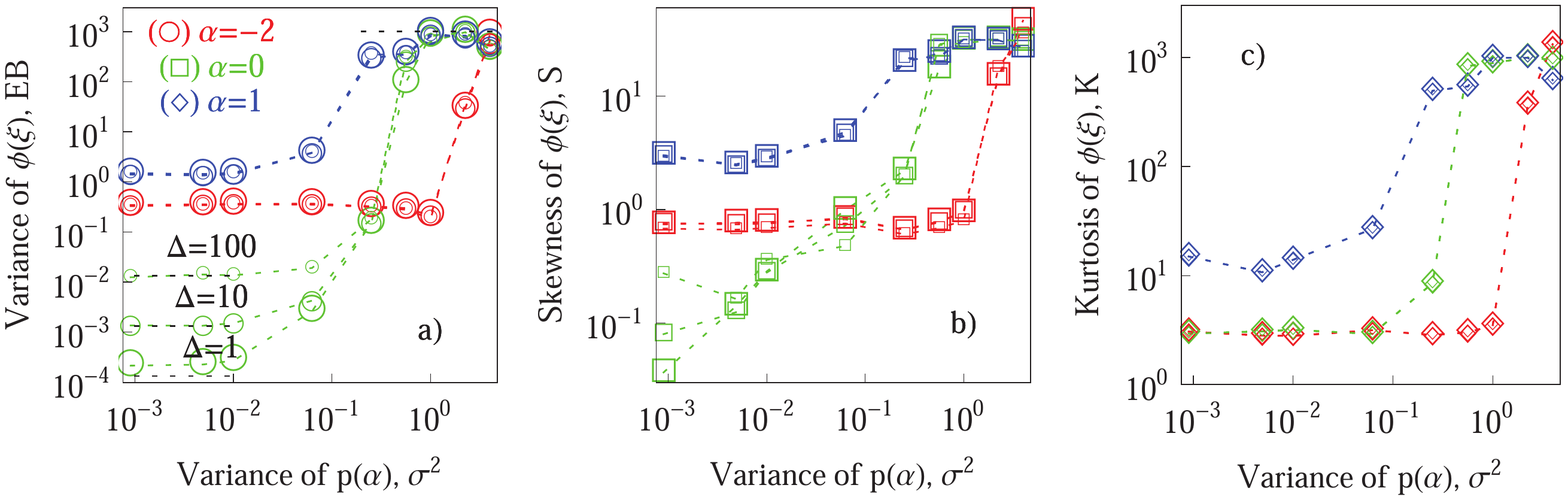}
\caption{Second, third, and fourth order moments of the amplitude scatter
distribution $\phi(\xi)$ for annealed noisy HDPs, computed for the parameters of
Fig.~\ref{fig-ann-tamsd}. Large, medium, and small symbols correspond to lag times
$\Delta=1$, $10$, and $100$, respectively. The dotted line in panel (a) at large
$\sigma^2$ is the ergodicity breaking parameter $\mathrm{EB}\approx N=10^3$,
indicative of the single trace dominance in this case, see
text. The dotted lines in panel (a) for small noise strength $\sigma^2$ stands for
$\mathrm{EB}_{\mathrm{BM}}(\Delta)$ given by Eq.~(\ref{eq-eb-bm}).}
\label{fig-ann-EB-higher}
\end{figure*}

\subsection{Noisy Brownian motion, $\alpha_0=0$}

For $\alpha_0=0$ and a small value $\sigma^2$ of the additional noise, as expected,
we observe small discrepancies from the canonical Brownian motion, as evidenced in
Fig.~\ref{fig-ann-tamsd}b. The behavior is ergodic, and the ergodicity breaking
parameter follows the known behavior (\ref{eq-eb-bm}) for Brownian motion, see
Fig.~\ref{fig-ann-EB-higher}. Most importantly, the ensemble averaged MSD equals the 
time averaged MSD, apart from very short lag times at which the relaxation from the 
initial value $x(0)=x_0$ occurs (compare Ref.~\cite{hdp2} for more details). At
longer lag times, the deteriorating statistics of the $\overline{\delta^2}$ traces
give rise to the typical cone-like scatter.

As the noise strength $\sigma^2$, the variance of the $\alpha$ distribution $p(
\alpha)$ is increased, see, for instance, in the panel for the noise strength
$\sigma=0.5$
in Fig.~\ref{fig-ann-tamsd}b: a more pronounced scatter of the $\overline{\delta
^2}$ traces emerges and, importantly, persists in the limit $\Delta/T\to0$. The
occurrence of progressively more distant particle excursions caused by
superdiffusive traces with $\alpha>0$ gives rise to a larger spread of the amplitude
scatter quantified by the distribution $\phi(\xi)$. The value of $\left<\overline{
\delta^2}\right>$ grows somewhat faster than the ensemble MSD (\ref{msd}) due
to these outliers, giving rise to larger values of the ergodicity breaking parameter
 $\mathrm{EB}$ (not shown). The time averaged MSD $\overline{\delta^2(\Delta)}$
scales linearly with the lag time $\Delta$, and, as they should, in the limit $\Delta
\to T$ the time averaged MSD settles back to the ensemble averaged MSD, due to the
pole in the definition (\ref{eq-TAMSD}) of the time average.

For even larger noise strength $\sigma^2$, the behavior of the time averaged MSD
and the ergodic properties are dominated by extreme events, that is, by single or
few trajectories in the data set with the largest exponent(s) yielding extremely
distant particle excursions. With an increasing width of the $\alpha$ distribution
$p(\alpha)$, the spread of the time averaged MSD grows, as well, as evidenced in
Fig.~\ref{fig-ann-tamsd}b. Similarly, for such large values of the noise
strength $\sigma^2$ the value of the ergodicity breaking parameter becomes
proportional to the number $N$ of recorded traces, witnessing the dominance of
single traces, each having the potential to be more extreme than the others,
compare Fig.~\ref{fig-ann-EB-higher}a. We refer the reader to Ref.~\cite{hdp2},
in which the critical properties of HDPs and the effects of the number of traces
are analyzed in the limit $\alpha_0\to2$.

For narrow distributions $p(\alpha)$ the spread $\phi(\xi)$ of individual
$\overline{\delta^2}$ traces is symmetric at short lag times $\Delta$, developing
a tail at larger lag times $\Delta$. This behavior can be rationalized in terms of
a generalized Gamma distribution (see Ref.~\cite{hdp}). The general features
of $\phi(\xi)$ are shown in Fig.~\ref{fig-ann-EB-higher} in terms of the higher
moments of this distribution. These are the skewness
\begin{equation}
S(\xi)=\frac{N^{-1}\sum_{i=1}^N(\xi-1)^3}{\left(N^{-1}\sum_{i=1}^N(\xi-1)^2\right)
^{3/2}}
\end{equation}
and the kurtosis 
\begin{equation}
K(\xi)=\frac{N^{-1}\sum_{i=1}^N(\xi-1)^4}{\left(N^{-1}\sum_{i=1}^N(\xi-1)^2\right)
^2},
\end{equation}
which complement the variance of $\phi(\xi)$ described by the ergodicity breaking
parameter (\ref{eq-eb-via-xi}). In Fig.~\ref{fig-ann-EB-higher}a we also observe
that for small noise strengths $\sigma^2$ the value $\mathrm{EB}(\Delta=1)$ for
noisy Brownian motion approaches $\mathrm{EB}_{\mathrm{BM}}(\Delta=1)$ given by
Eq.~(\ref{eq-eb-bm}), as expected. The values of the ergodicity breaking parameter
grow with $\Delta$, indicative of a bigger spread of the value $\overline{\delta^2}$
of individual traces (green points in Fig.~\ref{fig-ann-EB-higher}a).

\subsection{Subdiffusive noisy HDP, $\alpha_0=-2$}

For the subdiffusive case the time evolution of the ensemble and time averaged
MSDs is illustrated in Fig.~\ref{fig-ann-tamsd}a for different noise strengths
$\sigma^2$ of the $\alpha$ distribution. We observe that for the subdiffusive
value $\alpha_0=-2$ the same magnitude of the $\alpha$ variation causes a much
weaker effect as compared to the Brownian ($\alpha_0=0$) or superdiffusive ($\alpha
_0=1$) situations. The scatter of $\overline{\delta^2}$ remains nearly insensitive
to the lag time $\Delta$, similar to canonical HDPs \cite{hdp,hdp2}. The scaling of
the ensemble averaged MSD also agrees with that for HDPs \cite{hdp}. It is reached
after less than a dozen of steps during which the relaxation of the initial
condition occurs, compare Refs.~\cite{hdp,hdp2}. The scaling of the time averaged
MSD $\left<\overline{\delta^2}\right>$ remains linear and nearly unaffected
by changes of $\sigma^2$. The long time scaling of the MSD is also weakly sensitive
to $\sigma^2$ in the range considered here. 

Physically, for the subdiffusive case the spread of $\alpha_i$ should be $\gtrsim
\alpha_0$ to give rise to fast particle excursions (outliers). Thus, much larger
$\sigma^2$ values are required to disturb the spread of $\overline{\delta^2}$ for 
strongly subdiffusive noisy HDPs as compared to superdiffusive noisy HDPs shown in
Fig.~\ref{fig-ann-tamsd}c. This is our first important conclusion.

We rationalize the effects of the $\alpha$ spread further in terms of the width and
higher moments of the
amplitude scatter distribution $\phi(\xi)$. The results for $\alpha_0=0$, sub- and
superdiffusive annealed noisy HDPs are shown in Fig.~\ref{fig-ann-EB-higher}a. We
observe that all moments are typically smaller for the subdiffusive case reflecting
a less pronounced and asymmetric spread of $\overline{\delta^2}$. The skewness of
Brownian motion ($\sigma^2\to0$) tends to vanish, as it should, while for sub- and
superdiffusive noisy HDPs it attains finite values at $\sigma^2\to0$
(Fig.~\ref{fig-ann-EB-higher}b). This is due
to the inherent asymmetry of the $\phi(\xi)$ scatter even at $\sigma^2\to0$: it
features a tail at large $\xi$ values, a maximum at intermediate $\xi$, and vanishes
at $\xi\to0$ \cite{hdp}. Both skewness $S(\xi)$ and kurtosis $K(\xi)$ grow 
dramatically with $\sigma^2$ for all values of $\alpha_0$, as demonstrated in
Figs.~\ref{fig-ann-EB-higher}b,c. We checked that for $\sigma^2\to0$ the value
of the ergodicity breaking parameter in the limit $\Delta/T\ll1$ approaches that
for standard HDPs \cite{hdp}, as expected, while for a broad distribution of
$\alpha$ values the ergodicity breaking parameter increases and eventually
approaches the number of traces $N$ in the data set (single-trace domination), 
Fig.~\ref{fig-ann-EB-higher}a. The value of the ergodicity breaking parameter
$\mathrm{EB}$ for $\alpha_0<0$ is nearly unaffected by variations of $\sigma^2$
over a wide range, see the red symbols in Fig.~\ref{fig-ann-EB-higher}a. This
reflects the minor change in the spread of single traces $\overline{\delta^2}$
when $\sigma^2$ is varied, see Fig.~\ref{fig-ann-tamsd}a.

\subsection{Superdiffusive noisy HDPs, $\alpha_0=1$}
\label{sec-ann-super}

The ensemble and time averaged MSDs of superdiffusive noisy HDPs with $\alpha_0=1$
are shown in Fig.~\ref{fig-ann-tamsd}c. For small noise strengths $\sigma^2$ their
scaling agrees with the results for standard HDPs, Eqs.~(\ref{msd}) and
(\ref{eq-tamsd-hdp}). With increasing noise strength $\sigma^2$, the time averaged
MSD traces $\overline{\delta^2}$ grow dramatically, and for moderate and large lag
times $\Delta$ the time averaged MSD deviates progressively from the HDP scaling, 
that is ballistic for $\alpha_0=1$ (Fig.~\ref{fig-ann-tamsd}c). The scatter of the
individual time averaged MSDs $\overline{\delta^2}$ becomes progressively larger 
and asymmetric as the width of $p(\alpha)$ increases. The amplitude of the time
averaged MSD traces $\overline{\delta^2}$ for large values of $\sigma^2$ grows
significantly above the asymptote for undisturbed HDPs due to single trajectory
domination. Therefore, the moments of the scatter distribution $\phi(\xi)$
increase, see the blue symbols in Fig.~\ref{fig-ann-EB-higher}. For the later parts
of the trajectories, the ensemble averaged MSD increases very fast (see the right
panel in Fig.~\ref{fig-ann-tamsd}c) to meet the value of $\overline{\delta^2}$ in
the limit $\Delta=T$. For superdiffusive HDPs the moments of $\phi(\xi)$ are
larger than those for subdiffusive noisy HDPs with the same $\sigma^2$, 
compare the red and blue symbols in Fig.~\ref{fig-ann-EB-higher}.

\section{Noisy HDPs with quenched disorder}
\label{sec-quenched}

\begin{figure*}
\includegraphics[width=16cm]{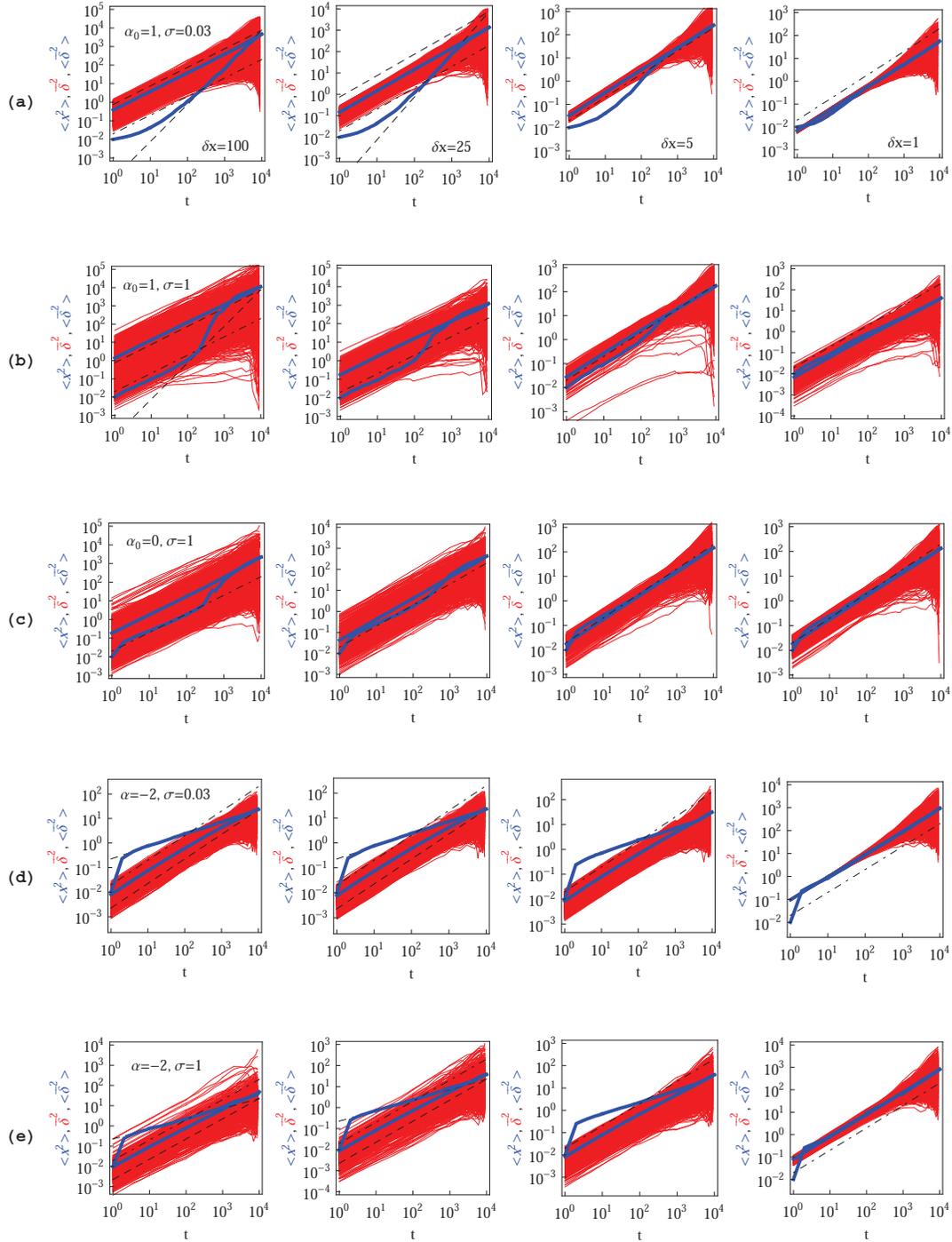}
\caption{Ensemble and time averaged MSDs and amplitude scatter of individual traces
$\overline{\delta^2}$ for noisy HDPs with quenched disorder. The values of $\alpha$,
$\sigma$, and domain size $\delta x$ are indicated in the plots. The panels (a) and
(b) are for subdiffusive noisy HDPs, panel (c) stands for noisy Brownian motion, 
and panels (d) and (e) represent superdiffusive noisy HDPs. The MSD is computed with
respect to the position of the center of the first domain, $\left<(x(t)-x_{c,1})^2
\right>$. Parameters: $T=10^4$, $N=10^3$, and $\delta x$ values are the same in each
column. The notations for the curves and asymptotes are the same as in
Fig.~\ref{fig-ann-tamsd}.}
\label{fig-quenched-tamsd-spread} 
\end{figure*}

We now turn to the situation of quenched disorder in a stratified environment, in
which evenly sized domains of width $\delta x$ have a diffusivity of the form
(\ref{eq-d-alfa}), centered within the domain, whose $\alpha$ value is noisy and
with distribution (\ref{eq-palfa}). In this quenched scenario, the particle
experiences the \emph{same\/} value of $\alpha$ each time it revisits a given
domain. The situation is illustrated in Fig.~\ref{fig-diff-coeff}c,d.

\subsection{Noisy Brownian motion, $\alpha_0=0$}

\begin{figure*}
\includegraphics[width=12cm]{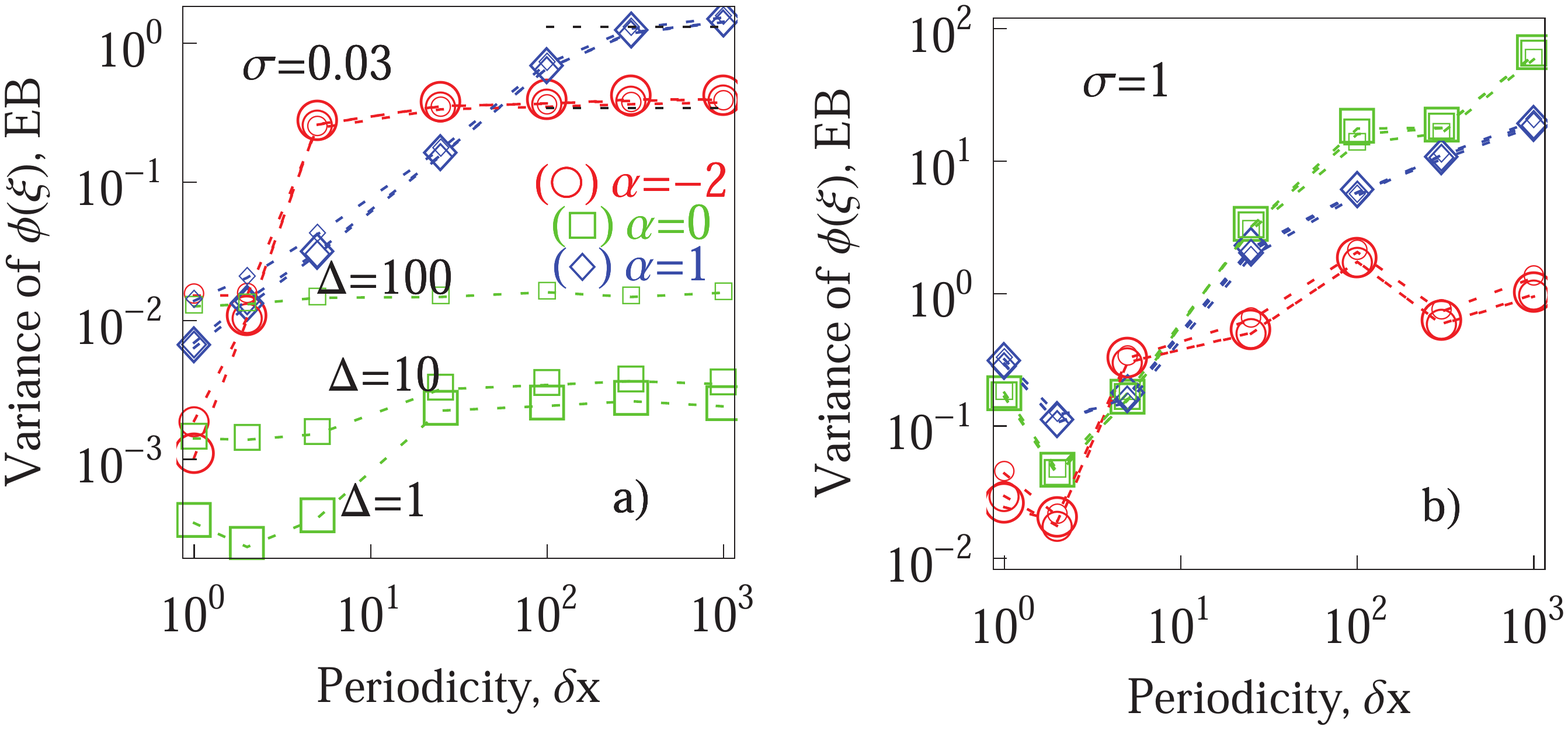}
\caption{Ergodicity breaking parameter of noisy HDPs with quenched disorder.
The parameters are the same as in Fig.~\ref{fig-quenched-tamsd-spread}, the
values for $\alpha$ and $\sigma$ being indicated in the plots. The black dotted
lines represent the ergodicity breaking parameter for the standard HDPs. The
large, medium, and small symbols correspond to the lag times $\Delta=1$, $10$,
and $100$, respectively.}
\label{fig-quenched-EB-higher}
\end{figure*}

For quenched, noisy Brownian motion we observe that for small noise strength
$\sigma^2$ the behavior, as expected, is very close to standard Brownian motion
(not shown). For a large value of $\sigma^2$, the spread of the amplitude of
individual time averaged MSDs $\overline{\delta^2(\Delta)}$ is non-negligible
even at short lag times $\Delta$, as shown in Fig.~\ref{fig-quenched-tamsd-spread}c.
This spread is more pronounced for larger periodicities $\delta x$ of the stratified
medium. For small $\sigma^2$ the ensemble averaged MSD $\left<(x-x_{c,1})^2\right>$
computed with respect to the center of the starting domain and the time averaged MSD
(thick blue lines) almost coincide for all lag times $\Delta$ (not shown).
Concurrently, the
ergodicity breaking parameter follows the Brownian asymptote (\ref{eq-eb-bm}), as
shown by the green symbols in Fig.~\ref{fig-quenched-EB-higher}a. For larger values
of the noise strength $\sigma^2$ the ergodicity breaking parameter deviates
pronouncedly from Eq.~(\ref{eq-eb-bm}) at short lag times $\Delta$, indicating the
occurrence of weak
ergodicity breaking, along with the disparity $\left<\overline{\delta^2}\right>\neq
\left<x^2\right>$, as witnessed by Fig.~\ref{fig-quenched-EB-higher}b. This
inequality is particularly pronounced for larger values of the noise strength
$\sigma^2$ and large periodicity $\delta x$, see the changes for varying $\delta x$
in Fig.~\ref{fig-quenched-tamsd-spread}c. For wider $\alpha$ distributions $p(
\alpha)$ the ensemble averaged MSD starts close to that of the asymptote for
standard Brownian motion, while at later times there occurs a crossover to the
curve for the time averaged MSD (left panel, Fig. \ref{fig-quenched-tamsd-spread}c.) 
This behavior is also typical for sub- and superdiffusive quenched noisy HDPs, 
see below. For $\sigma^2=1$ this transition occurs after $\sim10^3$ time steps
and becomes less pronounced for smaller periodicities $\delta x$ of the medium
(Fig.~\ref{fig-quenched-tamsd-spread}c).

\subsection{Superdiffusive noisy HDPs, $\alpha_0=1$}

In standard superdiffusive HDPs there exists a finite probability of particle
trapping in regions of low diffusivity near the origin, as witnessed by the
cusp around $x=0$ of the probability density function (\ref{eq-hdp-pdf}) \cite{hdp}. 
For noisy HDPs we find that for large values of the domain size $\delta x$ and small
noise strengths $\sigma^2$ the particle preferentially stays in the domain, in
which it was seeded, and the resulting ensemble averaged MSD is close to that of
the standard HDPs \cite{hdp,hdp2}. Here we again computed the MSD with respect to
the center $x_{c,1}$ of the seed domain in the form $x(t=0)-x_{c,1}=0.1$. The time
averaged MSD is equally close to the asymptote (\ref{eq-tamsd-hdp}) of the normal
HDP. Ensemble and time averaged MSDs converge at long lag times $\Delta\to T$, note
that the ensemble averaged MSD here is below the time averaged MSD, as evidenced by
Figs.~\ref{fig-quenched-tamsd-spread}a,b. 

We start with a narrow spread of $\alpha$ in the spatial domains corresponding to
$\sigma=0.03$. In this case we find that with decreasing domain
size $\delta x$ the amplitude scatter of individual time averaged MSDs shrinks and 
the amplitude of the trajectory mean $\left<\overline{\delta^2(\Delta)}\right>$ drops
substantially (Fig.~\ref{fig-quenched-tamsd-spread}a). The reason is that for a
small domain size there are almost no regions of fast diffusivity. For small values
of $\delta x$ the ensemble and time averaged MSDs converge and drop below the
Brownian asymptote, see the dashed-dotted line in the right graph in
Fig.~\ref{fig-quenched-tamsd-spread}a.
In such cases of smaller domain size the ergodicity breaking parameter attains
relatively small values, as shown in Fig.~\ref{fig-quenched-EB-higher}a, indicating
a more ergodic behavior. This effect of the noise is similar to that for noisy
CTRWs \cite{nCTRW}. As $\delta x$ increases, the ergodicity breaking parameter
approaches values close to those
of the standard HDP, $\mathrm{EB}(\Delta=1)\approx0.34$ for $\alpha_0=-2$ and 
$\mathrm{EB}(\Delta=1)\approx1.1$ for $\alpha_0=1$, with $T=10^4$ \cite{hdp,hdp2}. 
This is indicated by the dashed-dotted lines in Fig.~\ref{fig-quenched-EB-higher}a. 
Thus, frequent hopping events between individual domains destroys the characteristic
of the noise-free HDP scaling and causes the diffusion to be more ergodic. This is
our second important conclusion.

For larger $\sigma$ values the MSD stops following the HDP scaling law (\ref{msd})
and instead two nearly Brownian regimes are detected for short and long diffusion
times, see the left panel in Fig. \ref{fig-quenched-tamsd-spread}b. Similar to
noisy CTRWs \cite{nCTRW,pccp}, for noisy HDPs we observe a superposition of 
anomalous scaling for the MSD inherent to HDPs with the linear MSD increase due
to particle jumping between the stratified domains. The latter term contributes
stronger for smaller $\delta x$ values: after a given number of steps $T$ performed
the particle visits more $D(x)$ domains and its diffusion on the length scale $\gg
\delta x$ becomes effectively more normal and ergodic.

\begin{figure*}
\includegraphics[width=18cm]{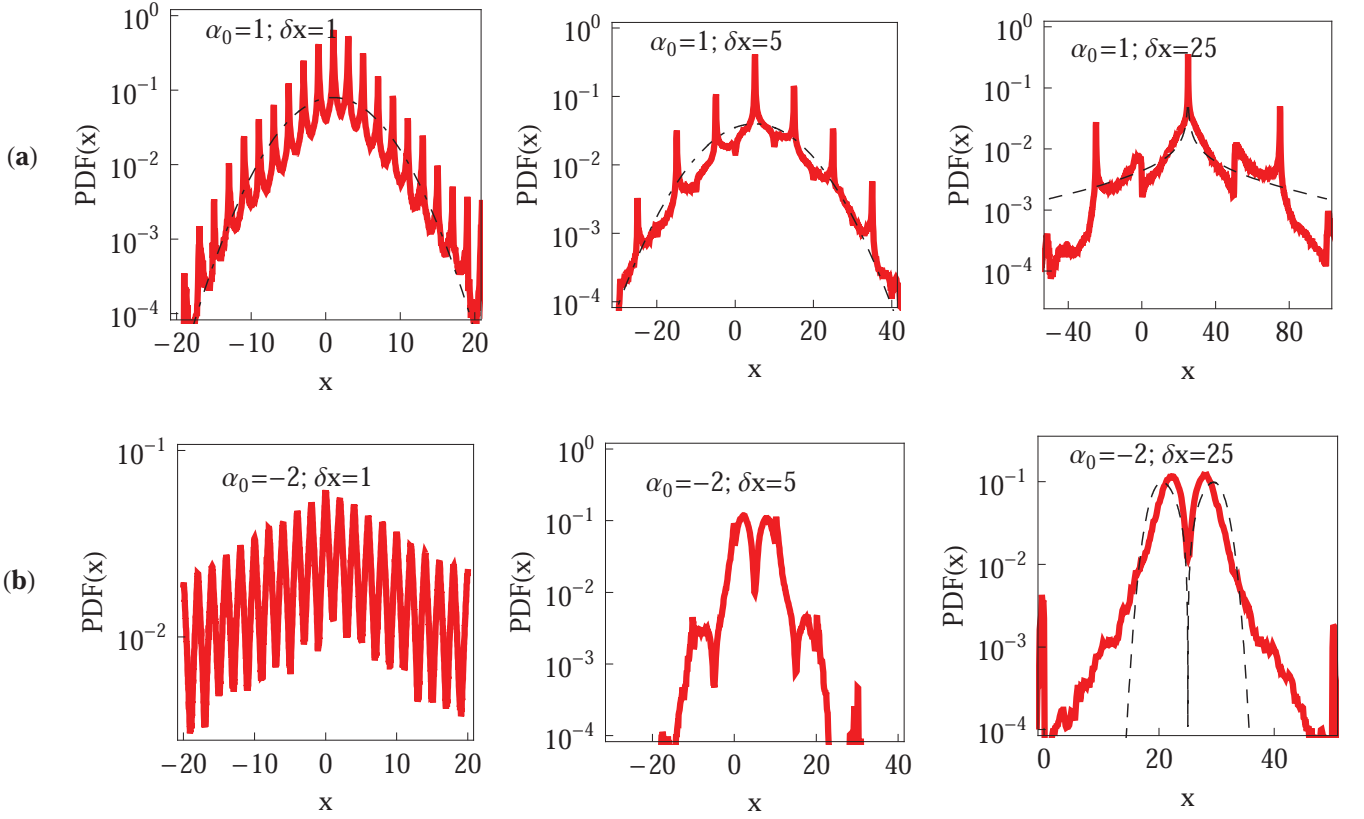}
\caption{Probability density functions of noisy HDPs with quenched disorder for
noise strength $\sigma^2$=1 and varying domain size $\delta x$. The other
parameters are the same as in Fig.~\ref{fig-quenched-tamsd-spread}. Panel (a)
represents superdiffusive noisy HDPs and panel (b) stands for subdiffusive noisy
HDPs. The shift of the peak positions with respect to those of the standard HDPs
\cite{hdp} is due to the shift $(\delta x)$ of the domain center positions,
see Eq.~(\ref{eq-xc-defin}). The dashed curves represent the probability density
functions of standard HDPs, given by Eq.~(\ref{eq-hdp-pdf}), while the dashed-dotted
curves are result (\ref{eq-brown-spread}).}
\label{fig-quenched-PDF}
\end{figure*}

The time averaged MSD is an approximately linear function of the lag time $\Delta$. 
For smaller domain size $\delta x$ we observe a more confined amplitude spread of
the time averaged MSD traces $\overline{\delta^2}$, see the evolution from left to
right in Figs.~\ref{fig-quenched-tamsd-spread}a,b. A
similar behavior occurs for subdiffusive noisy HDPs, as demonstrated in
Figs.~\ref{fig-quenched-tamsd-spread}d,e consistent with smaller values of the
ergodicity breaking parameter. This is our third main result. For
superdiffusive noisy HDPs, given large domain sizes $\delta x$, we observe more
distant particle excursions and thus a broader amplitude spread of individual time
traces $\overline{\delta^2(\Delta)}$, particularly for large values $\sigma^2$ of
the noise strength (Figs.~\ref{fig-quenched-tamsd-spread}e): at larger $\sigma^2$
we correspondingly obtain larger values of the ergodicity breaking parameter,
compare panels (a) and (b) in Fig.~\ref{fig-quenched-EB-higher}.

We find that the
distribution $\phi(\xi)$ of the amplitude scatter features a skewed form, which
is characterized by its second, third, and fourth moments corresponding to the
ergodicity breaking parameter, the skewness $S$, and the kurtosis $K$, respectively.
For larger values of $\sigma^2$, $S(\xi)$ and $K(\xi)$ grow with the domain size
$\delta x$ and are more irregular than the distribution $\phi(\xi)$ itself, due to
worsening statistics for higher order moments (not shown).
Note that for short lag times $\Delta/T\ll1$ the ergodicity breaking parameter for
large domain sizes $\delta x$ approaches the values of the corresponding normal
HDPs \cite{hdp}, compare Fig.~\ref{fig-quenched-EB-higher}a. At small domain size
$\delta x$ the non-ergodic properties of the standard HDPs are masked by the noise
in the stratified spatial domains. 

For superdiffusive HDPs the particles tend to localize in the center of each domain,
while for subdiffusive values $\alpha_0<0$ they tend to spread towards regions of
low diffusivity near the domain borders. In the long time limit the particles
spread over many domains, establishing the shape of the probability density function
$P(x,t)$ presented in Fig.~\ref{fig-quenched-PDF}. The local minima and maxima of
$P(x,t)$ correspond to the regions of low and fast diffusivity $D(x)$, respectively,
see Figs.~\ref{fig-diff-coeff}c,d and \ref{fig-quenched-PDF}. For relatively large
domain size $\delta x$ the probability density function of the noisy HDPs becomes
dominated by the contribution from the seed domain. The spreading of particles over
superdiffusive HDP domains in the long time limit is symmetric and nearly Gaussian,
\begin{equation}
P(x,t)=\frac{1}{\sqrt{4\pi D_{\mathrm{eff}}t}}\exp\left(-\frac{(x-x_{c,1})^2}{4
D_{\mathrm{eff}}t}\right)
\label{eq-brown-spread}
\end{equation}
with the effective diffusivity $D_{\mathrm{eff}}$. The mean particle displacement
with respect to the center of the seed domain vanishes, $\left<x(t\to\infty)\right>
\to0$. To compute $D_{\mathrm{eff}}$ analytically a homogenization procedure and
generic concepts of diffusion in random and highly heterogeneous media would need
to be applied \cite{math-heterogen}.

\subsection{Subdiffusive noisy HDPs, $\alpha_0=-2$}

Subdiffusive noisy HDPs in the quenched scenario share a number of trends with the
above descriptions of the cases $\alpha=0$ and $\alpha=1$. In particular, as the
domain size $\delta x$ decreases, the
amplitude spread of individual time averaged MSD traces $\overline{\delta^2}$
decreases (Fig.~\ref{fig-quenched-tamsd-spread}d,e). Because of the sublinear
scaling of the ensemble MSD of the normal subdiffusive HDPs ($\alpha_0<0$) the 
ensemble averaged MSD approaches the time averaged MSD $\left<\overline{\delta^2
(\Delta)}\right>$ from above. Moreover, the scaling of the ensemble averaged
MSD of subdiffusive noisy
HDPs with quenched disorder turns from subdiffusive to Brownian as the domain size
$\delta x$ decreases. The physical reason for this crossover behavior is the random
character of hops between domains with a varying local exponent $\alpha$. We
find that, similarly to superdiffusive noisy HDPs, the ensemble averaged MSD
initially follows the scaling (\ref{msd}) of normal HDPs while at later times a
nearly linear scaling is observed. For smaller periodicities $\delta x$ the linear
scaling becomes dominant, as demonstrated in Fig.~\ref{fig-quenched-tamsd-spread}e
from left to right. 

The probability density function of quenched noisy HDPs in the long time limit is
a combination of the superimposed local probability densities of the standard HDP.
For large periodicities $\delta x$ the probability density function is again 
dominated by the contribution from the seed domain, as can be seen in the right
panel of Fig.~\ref{fig-quenched-PDF}b. Similar to superdiffusive noisy HDPs we
find that the time averaged MSD is linear in the lag time, $\left<\overline{\delta
^2(\Delta)}\right>\sim\Delta$, while the amplitude spread of individual time
averaged MSDs grows with the noise strength $\sigma^2$ and becomes diminished 
for smaller medium periodicities $\delta x$. We also see that for subdiffusive
noisy HDPs the saturation of the ergodicity breaking parameter to the values of 
normal HDPs occurs at much smaller values of $\delta x$ as compared to 
superdiffusive noisy HDPs (Fig.~\ref{fig-quenched-EB-higher}a).

\section{Conclusions}
\label{sec-discussion}

We studied a stochastic process based on a combination of heterogeneous diffusion
processes with multiplicative noise and additional disorder of the environment,
distinguishing annealed and quenched scenarios. The environment was assumed to
be structured into periodic domains of given periodicity. We investigated the
diffusive and ergodic properties of these noisy heterogeneous diffusion
processes. The superposition of the additional stochasticity onto the standard
HDP with its deterministic variation of the diffusivity revealed a variety of new
features, the scaling relations for the ensemble and time averaged MSD of the
noisy HDPs being dramatically altered as compared to the normal HDP behavior.

For annealed disorder, the scaling exponent $\alpha$ of the diffusivity profile
switches in time and the gradient field of the particle diffusivity has a single
origin at $x=0$. We demonstrated how the Gaussian spread $p(\alpha)$ of the scaling
exponent gives rise to a strongly asymmetric scatter of individual time averaged MSD
traces. Rapidly switching diffusivity profiles in such an annealed environment
cause transient particle trapping in low-diffusivity regions. For superdiffusive
motion the effects of the $\alpha$ spread are more pronounced. In the case of a
quenched environment, a spatially stratified medium is modeled in terms of domains
of width $2\delta x$ with a normal distribution of the local HDP exponent. Upon
particle diffusion, the averaging is thus performed over ensembles of particle
trajectories generated for different spatial distributions of the scaling exponents
$\alpha$ in the domains. One of the key findings is that for small periodicity
$\delta x$ the sub- and superdiffusive scaling of normal HDPs cross over to a
linear growth of the ensemble averaged MSD as function of time. External noise
thus progressively masks the statistics of the underlying HDP.

What could be the physical phenomena captured by the noisy HDP discussed here?
From a biological perspective, the diffusion of small molecules in assemblies
of non-identical, interconnected cells is a relevant example. The cell-to-cell
variations of the diffusivity are inherent to biological tissues, while every 
individual cell features a space dependent diffusivity in its cytoplasm
\cite{lang11}. At cell-to-cell boundaries the diffusivity likely varies with a
jump, as captured by our stratified model of the quenched disorder, with possibly
discontinuous diffusivity across the system. We note that heterogeneous
diffusivities can, for instance, play a role in the formation of gradients of
morphogen molecules in a developing cell tissue \cite{karsten-morpho}, a process
known to involve features of anomalous diffusion. It also features a division of
fluxes of the molecules into fluxes through cells, across the outer cell membranes,
and transport in extracellular spaces \cite{berry15}.
Heterogeneous diffusion of water molecules in brain tissues
\cite{brain-directed-diffusion} and strongly heterogeneous structures of cardiac
muscle tissue with nontrivial cell-cell coupling \cite{cardiac14} could be another
example. Similarly, the domains
in the noisy HDP could represent internal compartments in a single cell.
The quenched case would correspond to static environments whereas the
annealed scenario would stand for environments, which change rapidly compared to
the typical crossing times between domains.

Our results for noisy HDPs could also be useful for the description of nano-objects
trapped in dynamical temperature fields \cite{cichos14} and of particles in strong
temperature gradients \cite{braun-thermo}. Another field of relevance is the tracer
diffusion in heterogeneous assemblies of distributed obstacles \cite{surya}
mimicking features of the cell cytoplasm \cite{lang11} and diffusion on chemically
and mesoscopically periodically patterned solid-liquid interfaces \cite{lindenberg}.
On a macroscopic scale, water diffusion in subsurface hydrology applications is to 
be mentioned \cite{hydro}, as well as tracer motion in porous heterogeneous media
\cite{porous-hetero}. For the latter there likely exists a distance-dependent
diffusivity within each pore
constructing a network governing the diffusion of water and contaminants in soil
specimen \cite{hydro}. Finally, in statistical models of financial stock price
variations \cite{finance-books} the terms stochastic versus correlated volatility
widely occur, representing the diffusivity in random walk models \cite{stanley05}. 
Some patterns of correlated or clustered volatility observed in financial data
thus correspond to a systematically varying diffusivity in our model of quenched
noisy HDPs. Some repeats of non-Brownian up-and-down trends in stock price
fluctuations \cite{stanley05}  can thus be considered as HDPs repeatedly
occurring in time.

\acknowledgments

We acknowledge funding from the Academy of Finland (FiDiPro scheme to RM) 
and the Deutsche Forschungsgemeinschaft (Grant CH 707/5-1 to AGC).

\end{document}